\documentclass[twocolumn,preprintnumbers,nofootinbib,prd,superscriptaddress,groupedaddress,aps]{revtex4-1}

\usepackage{graphicx,amssymb,amsmath,amsthm,amsfonts,epsfig,epsf}
\usepackage[outdir=./]{epstopdf}
\usepackage[linktocpage]{hyperref}
\usepackage[usenames]{color}
\usepackage{epstopdf}

\usepackage{aas_macros}
\usepackage{bm}
\usepackage{dcolumn}
\usepackage[latin1]{inputenc}
\usepackage{latexsym}
\usepackage{rotating}
\usepackage{color}
\usepackage{longtable}
\usepackage{enumerate}
\usepackage{tensor}
\usepackage{stmaryrd}
\usepackage{verbatim}

\usepackage{url}
\def\be{\begin{equation}}
\def\ee{\end{equation}}
\def\beq{\begin{eqnarray}}
\def\eeq{\end{eqnarray}}

\begin{document}

\title{On Nonsingular Solutions in Einstein-scalar-Gauss-Bonnet Cosmology}

\author{Laura Sberna}\email{lsberna@perimeterinstitute.ca}
\affiliation{Perimeter Institute for Theoretical Physics, 31 Caroline Street North Waterloo, Ontario N2L 2Y5, Canada.}
\author{Paolo Pani}\email{paolo.pani@roma1.infn.it}
\affiliation{Dipartimento di Fisica, ``Sapienza'' Universit\`a di Roma \& Sezione INFN Roma1, Piazzale Aldo Moro 5, 00185, Roma, Italy}

\begin{abstract} 
It is generically believed that higher-order curvature corrections to the Einstein-Hilbert action might cure the curvature singularities that plague general relativity. Here we consider Einstein-scalar-Gauss-Bonnet gravity, the only four-dimensional, ghost-free theory with quadratic curvature terms. For any choice of the coupling function and of the scalar potential, we show that the theory does not allow for bouncing solutions in the flat and open Friedmann universe. For the case of a closed universe, using a reverse-engineering method, we explicitly provide a bouncing solution which is nevertheless linearly unstable in the scalar gravitational sector. Moreover, we show that the expanding, singularity-free, early-time cosmologies allowed in the theory are unstable. These results rely only on analyticity and finiteness of cosmological variables at early times. 
\end{abstract}

\maketitle

\section{Introduction}
One of the main theoretical drawbacks of general relativity~(GR) is the rather generic formation of curvature singularities~\cite{HawkingEllis}, most notably in the interior of black holes and in the early universe.
It is commonly believed that such singularities should be resolved by higher-energy corrections to GR, which become important in the large-curvature regions close to the singularity. Higher-order curvature terms in the Einstein-Hilbert action are natural in low-energy effective string theories~\cite{Gross:1986mw} and in several modified theories of gravity~\cite{Clifton:2011jh,Berti:2015itd}.

Generally, higher curvature terms in the action lead to field equations of third (or higher) order, which are subject to Ostrogradsky's instability~\cite{Woodard:2006nt}. To quadratic order in the curvature, the only exception is the Gauss-Bonnet~(GB) combination. In four spacetime dimensions, the GB term is a total derivative and it would not contribute to the field equations, unless it is coupled to a scalar field, as in the so-called Einstein-scalar-Gauss-Bonnet~(EsGB) theory considered here.

Owing to the hypotheses of the singularity theorems~\cite{Penrose:1965i}, resolving curvature singularities requires the violation of some energy conditions for the effective stress-energy tensor of the theory, which is the case in EsGB gravity~\cite{Kanti:1998jd}. Indeed, EsGB gravity is well-known to admit nonsingular cosmological solutions~\cite{Kanti:1998jd,Rizos:1993rt,Antoniadis:1993jc,Easther:1996yd}. Recently, there has been some renewed interest in singularity-free GB cosmology~\cite{Kanti:2015dra,Kanti:2015pda}, although the nonsingular solutions known so far are unstable~\cite{Kawai:1998ab,Soda:1998tr,Kawai:1997mf}.

Similarly, the possibility of nonsingular \emph{bouncing} cosmologies is under intense study in several other modified theories of gravity~\cite{Nojiri:2017ncd}. For example, nonsingular bouncing solutions have been found in Galileon theories~\cite{Qiu:2011cy,Easson:2011zy,Ijjas:2016vtq,Ijjas:2016tpn}, in Horndeski theory~\cite{Rubakov:2014jja,Kobayashi:2016xpl,Akama:2017jsa}, as well as in $f(R)$ gravity~\cite{Odintsov:2015ynk} and others~\cite{Hendi:2016tiy}.
Recently, an effective-field-theory approach to nonsingular cosmologies has been developed in~\cite{Cai:2016thi,Cai:2017tku}

In these cases a classical bounce occurs at energy densities well below the Planck scale, and requires a violation of the null energy condition~(NEC). In addition, nonsingular quantum bounces were recently found in conformally lifted GR~\cite{Gielen:2015uaa}.
It is well-known that NEC violations might lead to various pathologies, e.g. to ghost and gradient instabilities and to Hamiltonian unbounded from below~\cite{Sawicki:2012pz}. Therefore, finding well motivated gravitational theories that allow for NEC violations while admitting viable bouncing cosmologies is of utmost importance.

The scope of this paper is to investigate bouncing and emergent  cosmological scenarios in EsGB gravity with a generic coupling function and a generic potential. 
Indeed, one may hope that --~for some specific choices of the coupling function and scalar potential~-- nonsingular, stable cosmologies exist. However, as we shall show, nonsingular emergent solutions are generically unstable, whereas nonsingular bouncing solutions exist only for a closed universe. While in the latter case we cannot provide a generic argument for the (in)stability, we explicitly construct some examples of nonsingular bouncing solutions which are nonetheless unstable.

A similar problem was recently studied~\cite{Kobayashi:2016xpl} for the more general Horndeski theory, which contains EsGB as a particular case. It was proved that --~in the case of a flat universe~-- no nonsingular solutions exist which are regular during the entire history of the universe, at least if the matter content during the entire evolution is described by a $k$-essence field~\cite{Kobayashi:2016xpl,Akama:2017jsa}. Our results are complementary to these because --~although restricted to EsGB gravity~-- they do not require any knowledge of the entire history of the universe nor on the matter content that might drive the evolution well after the nonsingular genesis or bounce. Furthermore, as we shall show, our results are also valid for the case of an open universe and for some particular couplings that evade the proof in Refs.~\cite{Kobayashi:2016xpl,Akama:2017jsa}.

\section{Einstein-scalar-Gauss-Bonnet Cosmology}
In the present work we study EsGB gravity, a gravitational model described by the following action (we use units such that $8\pi G=c=1$)
\begin{equation}\label{actionV} \resizebox{.43\textwidth}{!}{$\displaystyle
S=M_P^2\int \! \mathrm{d}^4x \sqrt{-g} \left[\dfrac{R}{2}-\dfrac{\partial_{\mu}\phi\partial^{\mu}\phi}{2}-V(\phi)+\dfrac{f(\phi)}{8} R^2_{\rm GB}\right],$}
\end{equation}
where $M_P$ is the Planck mass, $R_{\rm GB}^2=R_{\mu\nu\rho\sigma}R^{\mu\nu\rho\sigma}-4R_{\mu\nu}R^{\mu\nu}+R^2$ is the GB curvature invariant, $\phi=\phi(x^\mu)$ is a scalar field, and $f(\phi)$, $V(\phi)$ are generic functions of the scalar field. In our units, the scalar field $\phi$ is dimensionless, whereas $V$ and $f$ have dimensions of an inverse mass squared and of a mass squared, respectively. 
Since the GB term is a topological invariant, when $f(\phi)={\rm const}$ the theory reduces (modulo boundary terms) to Einstein theory minimally coupled to a self-interacting scalar field.

Other energy components, such as radiation, matter, dark energy, or other scalars that drive different stages of cosmic evolution are all contained in the matter action which should be included in Eq.~\eqref{actionV}. However, they are expected to be subdominant and hence negligible during the very early stages of the cosmological evolution. Since our conclusions on the viability of a nonsingular early universe in EsGB cosmology are negative, we will not be concerned with the otherwise essential question of how to connect that stage to the evolution at later times.

The field equations arising from minimization of action~\eqref{actionV} read
\begin{align}
\begin{cases}
&G_{\mu\nu}=T^{\text{eff}}_{\mu\nu},\\
&\dfrac{1}{\sqrt{-g}}\partial_{\mu}\left(\sqrt{-g}\partial^{\mu}\phi\right)=
V'(\phi)-\dfrac{f'(\phi)}{8}R^2_{\rm GB},
\end{cases}
\end{align}
where we defined the effective stress-energy tensor
\begin{equation}\label{teff}
T^{\text{eff}}_{\mu\nu}=\partial_{\mu}\phi{\partial}_{\nu}\phi
-\dfrac{1}{2}g_{\mu\nu}\left({\partial}\phi\right)^2-K_{\mu\nu}-g_{\mu\nu}V(\phi),
 \end{equation} 
with 
\begin{equation}
K_{\mu\nu}=\frac{g_{\mu\rho}g_{\nu\lambda}+g_{\mu\lambda}g_{\nu\rho}}{8\sqrt{-g}}
\epsilon^{\kappa\lambda\alpha\beta}\nabla_{\gamma}\left[\tilde{R}^{\rho\gamma}_{\alpha\beta}\partial_{\kappa}f(\phi)\right]\,,
\end{equation}
and $\tilde{R}^{\mu\nu}_{\rho\sigma}=\frac{1}{\sqrt{-g}}\epsilon^{\mu\nu\alpha\beta}R_{\alpha\beta\rho\sigma}$.

We consider a homogeneous and isotropic Friedmann-Robertson-Walker (FRW) universe, described by the following ansatz for the metric,
\begin{equation}
ds^2=-dt^2+a(t)^2\left( \frac{dr^2}{1-kr^2}+r^2d\Omega^2\right) \,,\label{metric}
\end{equation}
where $ k=0,-1,+1 $ for flat, open and closed universes, respectively. Consistently with the metric symmetries, we assume a homogeneous time-dependent scalar field, $\phi=\phi(t)$. 

The Friedmann equations in EsGB gravity read
\begin{eqnarray}\label{eqfV}
\begin{cases}
&\ddot{\phi}+3H\dot{\phi}-3f' \left(H^2+\frac{k}{a^2}\right)\left(H^2+\dot{H}\right)+V'(\phi)=0,\\
&3\left(1+\dot{f}H\right)\left(H^2+\frac{k}{a^2}\right)-\dfrac{\dot{\phi}^2}{2}-V(\phi)=0,\\
&2\left(1+\dot{f}H\right)\left(H^2+\dot{H}\right)+\\
&+\left(1+\ddot{f}\right)\left(H^2+\frac{k}{a^2}\right)=-\dfrac{\dot{\phi}^2}{2}+V(\phi),
\end{cases}
\end{eqnarray}
where $H=\frac{\dot{a}}{ a}$ is the Hubble parameter, a dot denotes a time derivative, while a prime stands for derivation with respect to $\phi$. As expected, the GB coupling enters only through derivatives of the coupling function $f(\phi)$, since the theory reduces to GR when $f(\phi)={\rm const}$.
\subsection{Violation of the energy conditions}\label{violation}
EsGB gravity has long been known to allow for violations of both the NEC and the strong energy conditions~(SEC). In the symmetry-reduced framework of FRW cosmology, both NEC and SEC involve the effective energy density $\rho\equiv T_{00}$ and the effective pressure $p\equiv T_{rr}/g_{rr}$.
Using the effective stress-energy tensor defined in Eq.~\eqref{teff}, and focusing for simplicity on the flat case ($ k=0 $), we can write the NEC as
\begin{equation}\label{NECv} 
(\rho+p)_{\rm GB}=2 H\frac{ A}{C}\geq0 \,,
\end{equation} 
and the SEC as the condition above supplemented by the further condition
\begin{equation}\label{SECv} 
(\rho+3p)_{\rm GB}=\frac{12 H  \dot{\phi} B }{C}\geq0 \,.
\end{equation}
In the above expressions, the functions $A$, $B$, and $C$ are defined by
\begin{align}\label{A}
\begin{split}
A=&12 H^3 \left(\dot{\phi}^4 f''-2 V+3\dot{\phi}^2\right)+12 H^2\dot{\phi} V'+\\
&+H \left(-12 V\dot{\phi}^2+4 V^2+5 \dot{\phi}^4\right)+36 H^5+\\
&-2 \dot{\phi} V' \left(2 V+\dot{\phi}^2\right) \,,
\end{split}
\end{align}
\begin{align}\label{B}
\begin{split}
B=&6 H^3 \dot{\phi} \left(\dot{\phi}^2 f''+4\right)+6 H^2 V'-12 H V\dot{\phi}+\\
&-V' \left(2 V+\dot{\phi}^2\right) \,, \\
\end{split}
\end{align}
\begin{align}\label{C}
\begin{split}
C=&-12 H^2 \left(2 V+\dot{\phi}^2\right)+36 H^4+12 V\dot{\phi}^2+\\
&+4 V^2+5 \dot{\phi}^4 \,.
\end{split}
\end{align}
It is straightforward to see that the above functions do not have a definite sign and, therefore, conditions~\eqref{NECv} and~\eqref{SECv} can be violated in EsGB gravity~\cite{Kanti:1998jd}.

\subsection{Perturbations of the FRW background}
\vspace{-5pt}
We are interested in the stability of the FRW universe in EsGB theory.
Following Refs.~\cite{Kawai:1998ab,Soda:1998tr,Kawai:1997mf}, we consider linear perturbations of the background metric~\eqref{metric} and of the scalar field. The former can be decomposed into scalar, vector, and tensor perturbations according to their behaviour under spatial coordinate transformations, whereas the latter transforms as a scalar quantity, $\phi=\phi_0(t)+\delta\phi(x^\mu)$ with $\delta\phi\ll\phi_0$. Below we discuss the full set of perturbations for the sake of completeness, but for our purpose it would be enough to focus on the scalar and tensor sectors.

\subsubsection{Scalar sector}
The scalar part of the metric perturbations couples to the perturbations of the scalar field; this is therefore the most involved sector. It is convenient to write the background and the perturbations in terms of conformal time $\tau$, defined by $dt=a(t)d\tau$. In this new coordinate, we decompose the scalar part of the perturbed metric in terms of four scalar functions ($\Phi$, $\Psi$, $B$ and $E$) as 
\begin{equation}
 \delta g_{\mu\nu}^S = a^2(\tau)\left(\begin{array}{cc}
                                    -2\Phi & B_{|i} \\
                                    B_{|j} & -2(\Psi \gamma_{ij}-E_{|ij} )
                                   \end{array}
\right)\,,
\end{equation}
where $\gamma_{ij}dx^idx^j=dr^2(1-kr^2)^{-1}+r^2d\Omega^2$ is the line element of the background three-geometry. The scalar field is also perturbed as $\phi(x)=\phi(\tau)+\delta \phi(x) $.  By choosing a gauge such that $E=B=0$, the gauge invariant variables (cf., e.g., Refs.~\cite{Bardeen:1980kt,Kodama:1985bj,Mukhanov:1990me}) reduce to $\Phi$, $\Psi$ and $\delta \phi$. 

We compute the scalar perturbation equations for a generic curvature $k$ and potential $V(\phi)$. The entire scalar sector reduces to four coupled ordinary differential equations, expressed here in physical time $t$,  
\begin{widetext}
	\begin{flushleft}
\begin{align}
a^6& \left(-\left(\delta\ddot{ \phi}+3 H \delta\dot{ \phi} +\delta \phi  \left(\kappa ^2+V''\right)+3 H^2 \ddot{\Psi } f'+6 H^3 \dot{\Psi} f'-3 \dot{\Psi} \dot{\phi}\right)\right)+a^4 \bigg(H^2 \left(3 \delta \phi  \dot{H} f''+\kappa ^2 \Phi  f'\right)+ \nonumber\\
&-2 \kappa ^2 \Psi  \dot{H} f'+H \left(4 \Phi  \dot{\phi}-6 \dot{\Psi } f' \left( k +\dot{H}\right)\right)-3 H^3 \dot{\Phi} f'+6 H^4 \Phi  f'-3 k  \ddot{\Psi} f'+2 \Phi \ddot{\phi}+\dot{\Phi} \dot{\phi}\bigg)+\nonumber \\
&+a^2 \left(3 k  \left(\delta \phi  \dot{H} f''+2 \Psi  \dot{H} f'-H \dot{\Phi} f'\right)+\Phi  f' \left(6 H^2 \left( k -2 \dot{H}\right)+k \kappa ^2 \right)\right)-6 k \Phi  \dot{H} f'=0 \,, \label{scalar1} \\
a^2& (3 a^2 H^3 \delta \dot{\phi} f'-a^2 \delta \dot{\phi} \dot{\phi}+\delta \phi \left(-a^2 \left(V'-3 H^3 \dot{\phi} f''\right)+\kappa ^2 \left(k +a^2 H^2\right)+
f'+3 k  H \dot{\phi} f''\right)+ \nonumber\\
&+2 \Psi  \left(3 k -\kappa ^2 a^2\right) \left(H \dot{\phi} f'+1\right)-9 a^2 H^2 \dot{\Psi}  \dot{\phi} f'-6 a^2 H \dot{\Psi}+3 k  H \delta \dot{\phi} f'-3 k  \dot{\Psi} \dot{\phi} f')+\nonumber\\
&+\Phi  \left(6 k -2 a^2 \left(3 H^3 \dot{\phi} f'+V\right)\right)=0\,,\label{scalar2}\\
a^2 &\left(-k  \left(\delta \dot{\phi} f'+\delta \phi \dot{\phi} f''\right)+H \left(k  \delta \phi f'+2 \Phi \right)+3 H^2 \Phi  \dot{\phi} f'\right)+\nonumber \\
&+a^4 \left(-H^2 \delta \dot{\phi} f'+\delta \phi \left(-H^2 \dot{\phi} f''+H^3 f'-\dot{\phi}\right)+2 H \dot{\Psi} \dot{\phi} f'+2 \dot{\Psi} \right)+k  \Phi \dot{\phi}f'=0 \,, \label{scalar3}\\
-a^2& \bigg(3 k  \delta \ddot{\phi} f'+6 k  \delta \dot{\phi} \dot{\phi} f''+6 H \delta \dot{\phi} \dot{H} f'+3 k  \delta \phi f^{(3)} \dot{\phi}^2+6 \delta \phi H \dot{H} \dot{\phi} f''+3 k  \delta \phi \ddot{\phi} f''+ \nonumber\\
&+2 \kappa ^2 \delta \phi \dot{H} f'-6 \dot{H} \dot{\Psi} \dot{\phi} f'-12 \Psi  \dot{H} \left(H \dot{\phi} f'+1\right)-9 H^2 \dot{\Phi} \dot{\phi} f'+\nonumber \\
&+\Phi  \left(2 \kappa ^2 H \dot{\phi} f'+18 H^3 \dot{\phi}  f'-6 H^2 \left(2 \dot{\phi}^2 f''+2 \ddot{\phi}  f'-1\right)+2 \kappa ^2-3 \dot{\phi}^2\right)-6 H \dot{\Phi}\bigg)+\nonumber\\
&+a^4 \bigg(+\Psi  \left(2 \kappa ^2 \dot{\phi}^2 f''+2 \kappa ^2 \ddot{\phi} f'+6 H^2 \left(\dot{\phi}^2 f''+\ddot{\phi} f'+1\right)+2 \kappa ^2-6 V+3 \dot{\phi} ^2\right)+\nonumber \\
&+3 \Big(-H^2 \delta \ddot{\phi} f'-\dot{\phi} \left(\delta \dot{\phi} \left(2 H^2 f''+1\right)-2 H f' \left(2 H \dot{\Psi}+\ddot{\Psi}\right)\right)+k  \dot{\Phi} \dot{\phi} f'+\nonumber \\
&+2 \Phi \left(\dot{H} \left(4 H \dot{\phi} f'+2\right)+k  \left(\dot{\phi}^2 f''-H \dot{\phi} f'+\ddot{\phi} f'\right)\right)+\nonumber \\
&+\delta \phi \left(V'-H^2 \left(f^{(3)} \dot{\phi}^2+\ddot{\phi} f''\right)+2 H \dot{\Psi} \dot{\phi}^2 f''+2 H \dot{\Psi} \ddot{\phi} f'+6 H \dot{\Psi}+2 \ddot{\Psi}\right)\Big)\bigg)=0 \label{scalar4}\,.
\end{align}
\end{flushleft}
\end{widetext}
With some abuse of notation, in the above equations we identify $\Phi$, $\Psi$ and $\delta \phi$ with the time-dependent part of the corresponding perturbation function, whereas $\kappa$ is the eigenvalue of the $ \nabla ^2 $ operator on $\gamma_{ij}$ and identifies the Fourier mode.
Equation~\eqref{scalar2} is an algebraic equation for $ \Phi(t) $ as a function of the other scalar perturbations, while the constraint, Eq.~\eqref{scalar3}, is to be imposed on the initial conditions. By substituting $ \Phi $ in Eqs.~\eqref{scalar1} and \eqref{scalar4}, we are left with a system of two ordinary differential equations for the perturbations $ \Psi $ and $ \delta \phi $. To the best of our knowledge, the above equations had never been derived before for generic $f(\phi)$, $V(\phi)$ and $k$.  

\subsubsection{Vector sector}
The vector part of the metric perturbations can be decomposed in terms of two vector functions ($S_i$ and $F_i$) as 
\begin{equation}
 \delta g_{\mu\nu}^V = a^2(\tau)\left(\begin{array}{cc}
                                    0 & -S_i\\
                                    -S_j &  F_{i|j}+F_{j|i}
                                   \end{array}
\right)\,,
\end{equation}
where ${S_i}^{|i}=0={F_i}^{|i}=0$, and the combination $P_i=S_i+a\dot F_i$ is gauge invariant. The perturbation equations for the vector sector do not depend explicitly on the potential, and thus take the same form as in Ref.~\cite{Kawai:1997mf}. Going back to physical time $t$, these equations read
\begin{align}
&\alpha (\frac{1}{2}\nabla ^2+k)P_i=0 \ ,\\
&\left[\alpha(P_{i\vert j}+P_{j\vert i})\right]_{,t}+2H\alpha (P_{i\vert j}+P_{j\vert i})=0 \ ,
\end{align}
where $\alpha=1+H\dot f$.

\subsubsection{Tensor sector}
Previous works~\cite{Kawai:1998ab,Soda:1998tr,Kawai:1997mf} have shown that early-universe GB cosmology is affected by an instability in the tensor sector. The tensor part of the metric can be decomposed in terms of a transverse and traceless tensor $h_{ij}$ as
\begin{equation}
 \delta g_{\mu\nu}^T = a^2(t)\left(\begin{array}{cc}
                                    0 & 0\\
                                    0 &  h_{ij}
                                   \end{array}
\right)\,.
\end{equation}
It is easy to show that, in Fourier space, each polarization mode satisfies the linear equation~\cite{Kawai:1997mf}
\begin{equation}
 \ddot h+\left(3H+\frac{\dot\alpha}{\alpha}\right)\dot h + c_s^2\left[\frac{\kappa^2-2k}{a^2}\right]h=0\,,
\end{equation}
where again $\alpha=1+H\dot f$ and we have defined the squared effective speed of sound,
\begin{equation}
 c_s^2(t)=\dfrac{1+\ddot{f}}{1+\dot{f} H}\,.\label{cs2}
\end{equation}
As expected, in the GR limit, $f(\phi)={\rm const}$ and $c_s^2(t)=1$. On the other hand, by using the background field equations of EsGB gravity, we can rewrite the above expression as
\begin{align}\label{scond}
c_s^2(t)=&-\left(2\frac{\dot{H}-{k }/{a^2}}{H^2+k/a^2}+5\right)+\nonumber\\
&+\frac{2 V(\phi)}{(1+\dot f H) \left(H^2+k/a^2\right)}\,.
\end{align} 
Absence of superluminal modes requires the effective speed of sound to be less than unity\footnote{However, it is worth mention that superluminality about a given cosmological solution does not necessarily imply causal paradoxes~\cite{Babichev:2007dw}.}, whereas stability requires $c_s^2>0$. Furthermore, the function $\alpha$ also appears in front of the kinetic term of $h_{ij}$ in the linearized action~\cite{Kawai:1998ab,Soda:1998tr}. Thus, absence of ghost instabilities requires $\alpha>0$. Overall, we must require
\begin{equation}
0\leq c_s^2(t) \leq 1, \quad \alpha(t) = 1+\dot{f} H >0 \,.
\end{equation}

For a free scalar field ($V(\phi)=0$) the first term in Eq.~\eqref{scond} is negative for nonsingular solutions at early times, and therefore such solutions are unstable~\cite{Kawai:1998ab,Soda:1998tr,Kawai:1997mf}. 
On the other hand, by introducing a potential $V(\phi)$, the second term in Eq.~\eqref{scond} can be positive and might cure the otherwise unavoidable instability of a solution approaching the nonsingular regime --~when energy conditions are violated. We will show, however, that the presence of the potential only allows for stable bounces in the case of closed universe ($k=1$). In the flat-universe case ($k=0$), bounces are forbidden by a no-go theorem that we prove below, and also emergent, nonsingular, solutions are strongly disfavoured. \\

For the purpose of studying the early-time behavior of cosmological solutions, we introduce the variables
\begin{equation}
g(t)=\dot{f}(\phi(t)) \,, \quad x(t)=\dot{\phi}(t) \,.
\end{equation}
Moreover, it is convenient to make use of Eq.~\eqref{scond} to write $V(\phi (t))$ as a function of $a(t)$, $g(t)$, $c_s^2(t)$, namely
\begin{align}
V=& \frac{1+g H}{2 } \bigg[\frac{3k}{a^2} +c_s^2 \left(\frac{k}{a^2} + H^2\right)+ \nonumber\\
&+\left(2 \dot{H}+5 H^2\right)\bigg]\,.
\end{align}
Finally, we can substitute the above expression into the system~\eqref{eqfV}; upon substitution, the Friedmann equations only contain the four dynamical variables
\begin{equation}\label{functions}
a(t), \quad g(t), \quad c_s^2(t), \quad x(t).
\end{equation} 
In the $k=0$ case, it is straightforward to show that the Friedmann equations do not depend explicitly on $ a(t) $, but rather on $ H(t) $.

\section{Cosmological bounce in EsGB gravity}
A bounce is a phase in the history of the universe connecting a contraction period with an expansion period.

In an homogeneous and isotropic cosmology, a bounce connects a phase where $\dot{a}<0$, $H<0$, to a phase where $\dot{a}>0$, $H>0$. Thus, a bounce requires $H$ crossing zero with a nonnegative derivative, $\dot{H}\geq 0$. 

In general, a bouncing cosmology might exist whenever one of the conditions of Penrose's singularity theorem~\cite{Penrose:1965i} fails. In the symmetry-reduced FRW cosmology, Friedmann equations reduce to
\begin{equation}
2\dot{H}=-\left( \rho+p \right)_{\rm GB} +2 \frac{k}{a^2}
\end{equation}
The above expression clearly shows that NEC violation, i.e. $\left( \rho+p \right)_{\rm GB} <0$, is a necessary condition for a bounce in the flat universe ($ k=0 $). When, on the other hand, the universe is closed ($k=1$), the only necessary condition for a bounce is the violation of the SEC~\cite{MolinaParis:1998tx,Novello:2008ra}.

As previously discussed, EsGB gravity allows for NEC and SEC violations. However, this does not imply that bouncing cosmological solutions necessarily exist (although some solutions have been found for specific ansatz~\cite{Bamba:2014zoa}), nor does the argument above inform us on the stability of such solutions. We discuss both issues in the following.

\subsection{Flat universe: a no-go theorem for bouncing solutions}
We now look for solutions to Eqs.~\eqref{eqfV} satisfying the following hypotheses:
\begin{enumerate}
	\item[(1)] solutions describe a cosmological bounce, i.e. $H(t_b)=0$ and $\dot H(t_b)\geq0$; 
	\item[(2)] solutions are stable (at least) with respect to tensor perturbations and the latter are subluminal, i.e. $0\leq  c_s ^2 (t) \leq 1$, at least around $t=t_{b}$;
	\item[(3)] all functions~\eqref{functions} are analytic around $t=t_{b}$.
\end{enumerate}
Owing to time translation invariance, we can set $t_{b}=0$ without loss of generality. 
We first expand around the bounce time,
\begin{align}
& H(t)=H_0+H_1 t +{\cal O}(t^2)\,,\\
& g(t)=g_0+g_1 t +{\cal O}(t^2)\,,\\
& c_s ^2 (t)=c_{s,0} ^2 +c_{s,1} ^2  t +{\cal O}(t^2)\,,\\
& x(t)=x_0+x_1  t +{\cal O}(t^2) \,.
\end{align}
The bounce condition requires $H_0=0$. We can now evaluate the cosmological equations~\eqref{eqfV} at first order in $t$. In particular, the constraint reads
\begin{equation}\label{O1}
H_1+\frac{1}{2} x_0 ^2+{\cal O}\left(t\right)=0
\end{equation}
This requires $H_1=\dot{H}(0)\leq 0$. Let us consider the case $\dot{H}(0)< 0$ first. This does not describe a bounce: instead, together with $H(0)=0$, describes a Big Crunch, i.e. a universe going from expansion to contraction. 
If, on the other hand, $\dot{H}(0)=0$, this also requires $x(0)=0$. We must now take into account higher-order terms in the $t$ expansion. The constraint equation at ${\cal O}(t^2)$ simply reads $H_2=\ddot{H}(0)=0$. The next order, ${\cal O}(t^3)$, imposes again $H_3+\frac{1}{2} x_1 ^2=0$, and the same argument applies to $H_3$. By generalizing to higher orders, we see that either the universe is undergoing a Big Crunch, $\dot H(0)<0$, or the solution is the uninteresting, static, one, i.e. $H(t)=0$.

The very same conclusion can be reached by expanding the scale factor $ a(t) $, rather than the Hubble parameter, around the bounce time. In other words, our argument only requires either $a(t)$ or $H(t)$ to be analytic around $t=t_b$. 

The above argument proves that no bounce solution exists in spatially flat, homogeneous and isotropic EsGB cosmology.
A similar result was recently obtained in~\cite{Kobayashi:2016xpl} for the more general Horndeski theory, which contains EsGB as a particular case.
The proof in~\cite{Kobayashi:2016xpl} excludes nonsingular solutions which are regular during the entire history of the universe and (possibly) in the presence of a $k$-essence field~\cite{Kobayashi:2016xpl,Akama:2017jsa}. On the one hand, our results are restricted to EsGB; we do not, on the other hand, make any assumption on the matter fields that might drive the evolution before or after the bounce.
It is also straightforward to show that our results apply also to the case of an open universe ($k=-1$). The case of a closed universe ($k=1$) is drastically different, as we shall now discuss.

\subsection{Bouncing solutions in a closed universe}  
\begin{figure}[th]
	\centering
	\includegraphics[width=0.40\textwidth]{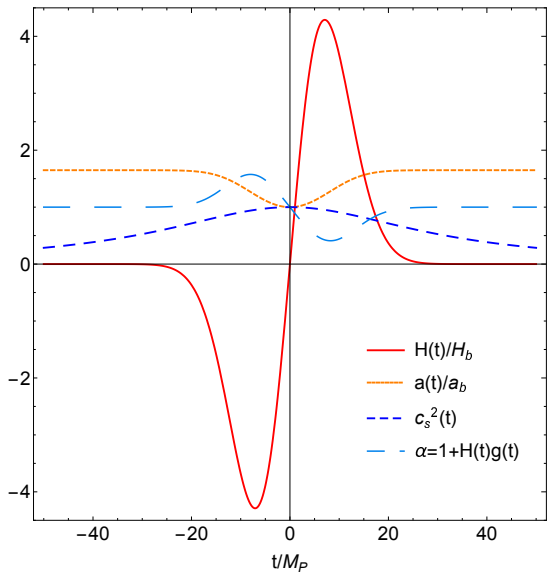}
	\caption{Our ansatz for the Hubble function and for the effective squared sound velocity of tensor perturbations. The parameters used for the plot are $t_b = 0$, $ G = 10^{-2}$, $ H_b = 10^{-2} $, $F = 10^{-3}  $, $ a_b=1 $. We also show the function $\alpha$ that appears in front of the kinetic term in the tensor sector. The condition $\alpha>0$ implies absence of ghosts. The equations of motion are integrated with initial condition $ g(t_i=-50)=1 $.}\label{aHcs}
\end{figure}

When $ k=1$, the argument outlined in the previous section does not apply. Indeed, in this case the theory allows for a classical cosmological bounce with stable tensor perturbations. 
We show this explicitly through a reverse-engineering method (cf., e.g., Refs.~\cite{Ijjas:2016vtq,Ijjas:2016tpn}). We look for solutions of the system~\eqref{eqfV} for which the Hubble function reads
\begin{equation}\label{Hbounce}
H(t)=H_b \, (t-t_b) \, e^{- G (t-t_b)^2} \,,
\end{equation}
whereas the squared speed of sound reads
\begin{equation}\label{cbounce}
c_s^2(t)=\frac{1}{1+F \, (t-t_b)^2} \,.
\end{equation}
In the equations above, $ H_b $, $ F $, and $ G $ are arbitrary, positive constants. This cosmological ansatz is shown in Fig.~\ref{aHcs}. By construction, Eqs.~\eqref{Hbounce},~\eqref{cbounce} describe a stable (in the tensor sector) universe going into a bounce at $ t_b $. First-order tensor perturbations display a maximum sound velocity at the bounce, $c_s^2(t_b)=1$, and are stable throughout the evolution. Furthermore, one can require the evolution to be classical --~i.e. the gravitational energy throughout the bounce to be much lower than the Planck energy~-- by choosing an appropriate value for $ H_b $, the Hubble parameter at the bounce, and $ G $. Indeed, the maximum gravitational energy density reached during the bounce reads $ H^2_{max}=\frac{H^2_b}{2 e G} $.
%

\begin{figure}[!t]
	\centering
	\includegraphics[width=0.40\textwidth]{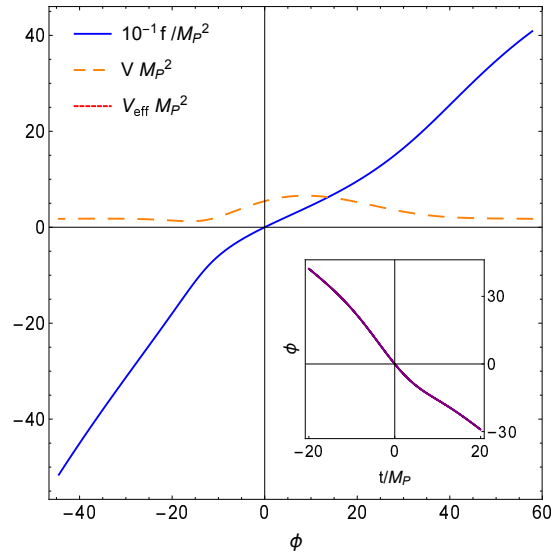}
	\caption{Potential $V$, coupling function $f$, and effective potential $V_{\rm eff}$ as a function of the scalar field for the bouncing solution shown in Fig.~\ref{aHcs}. The inset shows the corresponding scalar field as a function of time. Both the scalar field and the coupling function $ f $ are defined up to an integration constant (here $ \phi(t_b)=0 $, $ f(t_b)=0 $).}\label{Vf}
\end{figure}

By imposing the ansatz above to be a solution of the cosmological equations of motion, one obtains the functions $ x(t) $ and $ g(t) $. Further integration in time yields the scalar field as a function of time, as well as the scalar potential and coupling as functions of the scalar field. The result is shown in Fig.~\ref{Vf}: both $V(\phi)$ and $f(\phi)$ are smooth, simple, and single-valued. 

The potential $V$ and the effective potential $ V_{\rm eff}=V -\left( f/8 \right)R_{\rm GB} $ both display a maximum in the vicinity of $\phi_b=\phi(t_b)$, whereas $f(\phi)$ is a simple linear function near the bounce. Note that the scalar field diverges at $t\to\pm\infty$ but, strictly speaking, our solution needs to be valid only near $t\approx t_b$. Indeed, for $t\ll t_b$ and for $t\gg t_b$ this solution should be smoothly connected to a different evolution which is governed by other forms of matter fields, neglected here.
Finally, we checked that this solution satisfies the no-ghost condition.

\begin{figure}[!t]
	\centering
	\includegraphics[width=0.40\textwidth]{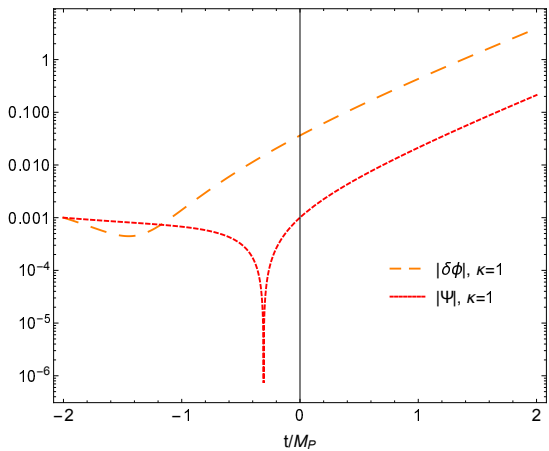}
	\caption{Example of two scalar perturbations (mode $ \kappa=1$) on the bouncing solution shown in Figs.~\ref{aHcs} and~\ref{Vf}. Both $\Psi$ and $\delta \phi$ are clearly unstable. The initial conditions for numerical integration are $\delta \phi (t_i)= 10^{-3}$, $\Psi(t_i)=10^{-3}$ and $\dot{\Psi}(t_i)=-10^{-3} $ at $ t_i=-2 M_P$.  
	}\label{ScalPert}
\end{figure}

While the tensor sector of the perturbations is, by construction, well-behaved for this solution, we can easily expect the scalar perturbations to be unstable, as a result of the maximum displayed by the potential and the effective potential, cf.\ Fig.~\ref{Vf}. Indeed, by numerically integrating Eqs.~\eqref{scalar1}--\eqref{scalar3} for this background configuration, we have confirmed this expectation. A representative example of scalar perturbations is shown in Fig.~\ref{ScalPert}:  even if we impose the scalar perturbations to be small close to the bounce, the nonsingular phase causes them to grow and eventually become as large as the background, i.e. ${\cal O}(1)$ or larger.

A similar result is obtained when choosing different bouncing ansatz for the Hubble function and for the speed of sound. Thus,  although we cannot find a general argument for the instability, we believe that this is a quite generic feature of bouncing solutions in EsGB cosmology: even when the instability is absent in the tensor sector, it re-emerges in the scalar perturbations.

\section{No stable emergent nonsingular universe in EsGB gravity}
We now turn to the discussion of the existence of nonsingular emergent, stable solutions in EsGB isotropic cosmology.  We focus, for simplicity, on flat FRW solutions ($ k=0 $).

We look for a cosmological solution satisfying the following assumptions:
\begin{enumerate}
	\item[(I)] the solution describes a nonsingular expanding universe, i.e. $H(t)\geq 0$ at early times;
	\item[(II)] similarly to the aforementioned bouncing solutions, we require stability with respect to (at least) tensor perturbations and subluminality, i.e. $0\leq  c_s ^2 (t) \leq 1$, $\forall t$;
	\item[(III)] all dynamical variables~\eqref{functions} are analytic in the infinite past,
	\begin{align}
	& H(t)= \sum_{n=0}^{+\infty}\dfrac{H_n}{t^n} = H_0+H_1 \dfrac{1}{t} +O(\dfrac{1}{t^2});\\
	& g(t)=\sum_{n=0}^{+\infty}\dfrac{g_n}{t^n} =g_0+g_1 \dfrac{1}{t} +O(\dfrac{1}{t^2});\\
	& c_s ^2 (t)=\sum_{n=0}^{+\infty}\dfrac{c^2 _{s,n}}{t^n} =c_{s,0} ^2 +c_{s,1} ^2 \dfrac{1}{t} +O(\dfrac{1}{t^2});\\
	& x(t)=\sum_{n=0}^{+\infty}\dfrac{x_n}{t^n} =x_0+x_1 \dfrac{1}{t} +O(\dfrac{1}{t^2}) \,;
	\end{align}
	\item[(IV)] the scalar field is regular in the infinite past. \label{assV}
\end{enumerate}
Condition~(III) implies $ \phi (t)\sim x_0 t +x_1\log|t|+... $ in the infinite past. Thus, regularity demands $ x_0=0=x_1 $, although the condition $x_0=0$ is sufficient for our purpose.
The asymptotic expansion of the field equations in the limit $ t\rightarrow-\infty $ imposes
\begin{equation}\label{key}
H_0= \frac{1 - c_{s,0} ^2}{c_{s,0} ^2\, g_0}\,,
\end{equation}
so that the condition $H_0\geq0$ is satisfied when $g_0>0$ (and $0\leq c_s^2\leq 1$ as requested by stability). Furthermore,
\begin{equation}\label{key2}
x_0=\frac{(1 - c_{s,0} ^2)^{3/2}}{c_{s,0} ^{3/2} \, g_0} \ .
\end{equation}
Therefore, regularity of the scalar field ($x_0=0$) implies $c_{s,0} ^2=1  $, which saturates the stability requirement. As a result, $ H_0=0 $. The constraint additionally requires
\begin{equation}\label{key3}
\frac{2H_1-x_1^2}{H_1}=0 \ ,
\end{equation}
which implies $H_1=x_1^2/2>0$ (strictly positive, since the constraint is singular when $H_1=0$). Therefore, $ H(t)\sim H_1/ t $ is negative in the infinite past, i.e., the Hubble function describes a contracting rather than an expanding universe, which is in conflict with assumption~(I). 

The argument outlined above proves that EsGB gravity, under the assumptions (I-IV), does not admit a stable, nonsingular and expanding solution. 
Our argument relies on the fact that $V(\phi)$, $g(\phi)$ and $\phi$ do not blow up at $t \rightarrow - \infty $. While a linearly divergent (in time) scalar field might not necessarily imply a divergence of the curvature, such configuration is usually prone to other types of instabilities~\cite{Ogawa:2015pea,Takahashi:2016dnv}. We also note that we did not make use of the condition $x_1=0$, i.e., our no-go argument includes the case in which the scalar field diverges logarithmically at early times.

Similarly to the no-go result for bouncing solutions previously derived, also in this case our results are complementary to those of Refs.~\cite{Kobayashi:2016xpl,Akama:2017jsa}, although restricted to the EsGB case.
Besides not making any assumption on the matter content at late times, our argument also applies to the case in which $c_s^2\to0$ at $t\to-\infty$, which is a condition that evades the no-go theorem of Refs.~\cite{Kobayashi:2016xpl,Akama:2017jsa}. Indeed, the theorem can be evaded if the function $F_T$ defined in Ref.~\cite{Kobayashi:2016xpl} vanishes sufficiently fast at past infinity. In our notation, $F_T\equiv \alpha c_s^2 $. Since $\alpha\neq0$ to avoid strong coupling, the condition $c_s^2\to0$ is equivalent to $F_T\to0$.

\begin{figure}[!ht]
		\centering
		\includegraphics[width=0.4\textwidth]{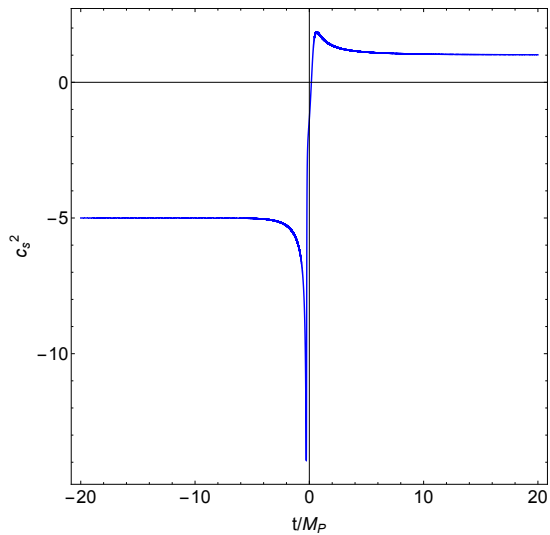}
		\caption{The squared sound velocity $c^2 _s$ of tensor perturbations of an emergent solution, as a function of rescaled time~$ t\rightarrow t/\sqrt{\vert \lambda\vert} $, in EsGB gravity with negative quadratic coupling $f(\phi)=\lambda \phi ^2$, $ \lambda <0 $ and no potential, $ V=0 $ (cf.\ Ref.~\cite{Sberna:2017nzp} for more details). This nonsingular solution is asymptotically flat in the infinite past.}\label{phi2-csk_0d_-1}
	\end{figure}

Finally, the above procedure also helps understanding the instability of nonsingular emergent solutions previously studied in the literature. The latter were found in the absence of scalar potential, $V(\phi)=0$. From condition~(III), the scalar potential in the infinite past reads
\begin{equation}
V(t)=\dfrac{1}{2}\left( 1+g_0 H_0 \right) \left( 5 +c^2 _{s,0} \right) H_0 ^2 + O(t^{-1})=0 \,,
\end{equation}
and it vanishes only if either $H_0 = - {1}/{g_0}$, $H_0=0$, or $c^2 _{s,0} = -5$. It is straightforward to see that the first case does not satisfy the field equations, whereas the second case leads again to a contracting universe that does not satisfy condition~(I) above. Finally, condition $c^2 _{s,0} = -5$ implies an instability.  Figure~\ref{phi2-csk_0d_-1} shows an example of a emergent solution satisfying this property. This solution was first found in Ref.~\cite{Antoniadis:1993jc} (cf. also Ref.~\cite{Sberna:2017nzp}), and its stability was studied in Ref.~\cite{Kawai:1997mf}. It is interesting that our argument provides a simple, analytical, confirmation of the instability. The same argument generalizes straightforwardly to the more generic case in which the potential is zero only at early times, $V(t\to\infty)=0$, but is otherwise nonvanishing. 

\section{Conclusion}

We have studied bouncing and emergent nonsingular cosmological solutions in EsGB gravity with a generic coupling function and generic scalar potential. 
Our results are twofold. On the one hand, under rather generic hypotheses (most notably analyticity of the dynamical variables), we have shown that EsGB does not admit bouncing solutions in the case of flat and open universe, and it does not admit stable emergent solutions in flat FRW universe (except for the trivial case of de Sitter-like solutions, which are anyway geodesically incomplete in the past~\cite{Borde:2003}, and thus cannot describe a nonsingular universe). 
These results are complementary to those recently obtained in~\cite{Kobayashi:2016xpl} for a flat universe in the more general Horndeski theory. Although restricted to EsGB theory, our analysis does not contain any assumption on the matter fields that might drive the evolution before or after the bounce or after the nonsingular genesis. 

On the other hand, we have shown that EsGB gravity admits nonsingular bouncing solutions in the case of a closed universe ($k=1$), and that these can be stable under tensor perturbations. The existence of similar solutions was recently proven for Galileon theories~\cite{Ijjas:2016vtq,Ijjas:2016tpn}. At variance with the latter case, our solutions require $k=1$, but the theory is considerably simpler, since it depends only on the scalar potential and on the single GB coupling function.

Unfortunately, these bouncing solutions are plagued by an instability in a sector that was previously believed to be pathology-free in GB cosmology, namely the scalar gravitational sector. Even though we could not prove in general that such instability is unavoidable, we believe that our result is pointing at a new obstacle in GB cosmology, and that (at the very least) fine-tuning of the arbitrary functions will be required in order to overcome it.
Finally, even assuming this instability can be somehow circumvented, a bouncing solution in closed FRW universe will likely require a subsequent inflationary phase, in order to match cosmological observations, which have found no sign of spatial curvature so far~\cite{Ade:2015xua}.

Overall, our results strongly suggest that no viable, singularity-free cosmological solution exists in EsGB gravity. It would be interesting to investigate whether this result can be circumvented by generalizing action~\eqref{actionV}, for example by adding a nonminimal scalar coupling to the Ricci scalar or to the kinetic term. Due to the presence of the GB terms, these couplings cannot be reabsorbed trivially by a conformal transformation or field redefinition, and have to be studied separately. Likewise, it would be interesting to extend our analysis to the more general Horndeski case.

\begin{acknowledgments}
We are indebted to Yong Cai, Tsutomu Kobayashi and Alexander Vikman for useful correspondence and discussions.
Research at Perimeter Institute is supported by the Government of Canada through Industry Canada and by the Province of Ontario through the Ministry of Economic Development $\&$ Innovation. 
\end{acknowledgments}
%
%
\bibliographystyle{utphys}
\bibliography{Ref}
\end{document}